
\input phyzzx
%
%
\def\abstract{\vskip\frontpageskip
  \centerline{\bf {\titlestyle Abstract}} \vskip\headskip}

\def\refout{\par \penalty-400 \vskip\chapterskip
  \spacecheck\referenceminspace \immediate\closeout\referencewrite
  \referenceopenfalse
  \line{\hfil {\titlestyle {\bf References}} \hfil}\vskip\headskip
  \input \jobname.refs}

\def\chapter#1{\par \penalty-300 \vskip\chapterskip
  \spacecheck\chapterminspace
  \chapterreset \titlestyle{{\bf \chapterlabel \ \ #1}}
  \nobreak\vskip\headskip \penalty 30000
  \wlog{\string\chapter\ \chapterlabel}}
%
%
\def\NPrefs{\let\therefmark=\NPrefmark \let\therefitem=\NPrefitem}
\def\NPrefmark#1{[#1]}
\def\NPrefitem#1{\refitem{[#1]}}
\NPrefs
%
%
\def\e{\, {\rm e}}
\def\Z{{\bf Z}}
\def\R{{\bf R}}
%
%
%
\REFSCON\dika{J. Distler and H. Kawai,
           {\it Nucl.\ Phys.\ }{\bf B321} (1989) 509;
           J. Distler, Z. Hlousek and H. Kawai,
           {\it Int.\ J. of Mod.\ Phys.\ }{\bf A5} (1990) 391; 1093;
           F. David, {\it Mod.\ Phys.\ Lett.\ }{\bf A3} (1989) 1651.}
\REFSCON\seirev{N. Seiberg,
           {\it Prog.\ Theor.\ Phys.\ Suppl.\ }{\bf 102} (1990) 319.}
\REFSCON\gouli{M. Goulian and M. Li,
           {\it Phys.\ Rev.\ Lett.\ }{\bf 66} (1991) 2051.}
\REFSCON\dfku{P. Di Francesco and D. Kutasov,
          {\it Phys.\ Lett.\ }{\bf 261B} (1991) 385;
           Y. Kitazawa, {\it Phys.\ Lett.\ }{\bf 265B} (1991) 262;
           N. Sakai and Y. Tanii,
           {\it Prog.\ Theor.\ Phys.\ }{\bf 86} (1991) 547;
           V.S. Dotsenko, Paris preprint PAR--LPTHE 91--18 (1991).}
\REFSCON\polch{J. Polchinski,
           {\it Nucl. Phys.\ }{\bf B324} (1989) 123;
           {\it Nucl.\ Phys.\ }{\bf B346} (1990) 253.}
\REFSCON\grklne{D.J. Gross, I.R. Klebanov and M.J. Newman,
           {\it Nucl.\ Phys.\ }{\bf B350} (1991) 621.}
\REFSCON\polyakov{A.M. Polyakov,
           {\it Mod.\ Phys.\ Lett.\ }{\bf A6} (1991) 635.}
\REFSCON\satafact{N. Sakai and Y. Tanii,
           Tokyo Inst.\ of Tech.\ and Saitama preprint
           TIT/HEP--173, STUPP--91--120 (1991);
           TIT/HEP--179, STUPP--91--122 (1991).}
\REFSCON\witten{E. Witten, Princeton preprint IASSNS--HEP--91/51 (1991).}
\REFSCON\klpo{I.R. Klebanov and A.M. Polyakov,
           {\it Mod.\ Phys.\ Lett.\ }{\bf A6} (1991) 3273.}
\REFSCON\masata{Y. Matsumura, N. Sakai and Y. Tanii,
           Tokyo Inst.\ of Tech.\ and Saitama preprint
           TIT/HEP--186, STUPP--92--124 (1992).}
\REFSCON\beku{M. Bershadsky and D. Kutasov, Princeton and Harvard
           preprint PUPT--1283, HUTP--91/A047 (1991);
           Y. Tanii and S. Yamaguchi, Saitama preprint STUPP--91--121
           (1991), to appear in {\it Mod.\ Phys.\ Lett.\ }{\bf A}.}
\REFSCON\gool{P. Goddard and D. Olive,
           in {\it Vertex Operators in Mathematics and Physics},
           eds.\ J. Lepowsky et al., (Springer, Heidelberg, 1985) p.\ 51;
           {\it Int.\ J. of Mod.\ Phys.\ }{\bf A1} (1986) 303.}
\REFSCON\rose{M.E. Rose, {\it Elementary Theory of Angular Momentum}
           (John Wiley \& Sons, New York, 1957).}
\refsend
%
%
\unnumberedchapters
\Pubnum={TIT/HEP--187 \cr STUPP--92--125}       
\titlepage
\title{\bf Interaction of Tachyons and Discrete States
in $c \! = \! 1$ 2-D Quantum Gravity}
\author{Yoichiro Matsumura, Norisuke Sakai}
\address{Department of Physics, Tokyo Institute of Technology \break
         Oh-okayama, Meguro, Tokyo 152, Japan}
\andauthor{Yoshiaki Tanii}
\address{Physics Department, Saitama University \break
         Urawa, Saitama 338, Japan}
\abstract
The two-dimensional (2-D)
quantum gravity coupled to the conformal matter with $c=1$ is studied.
We obtain all the three point
couplings involving tachyons and/or discrete states via operator
product expansion.
We find that cocycle factors are necessary and construct them
explicitly.
We obtain an effective action for these three point couplings.
This is a brief summary of our study of couplings of tachyons and
discrete states, reported at the workshop in Tokyo Metropolitan
University, December 4-6, 1991.
\endpage
%
%
Recently the understanding of two-dimensional (2-D) quantum gravity
has advanced significantly.
There are two main motivations to study the 2-D quantum
gravity coupled to matter.
Firstly,  it is precisely a string theory when the 2-D
space-time is regarded as the world sheet for the string.
Secondly, it provides a toy model for the quantum gravity in four
dimensions.
There are two approaches to study the 2-D quantum gravity:
the matrix model as a discretized theory and the Liouville theory
as a continuum theory \NPrefmark{\dika, \seirev}.
The former can provide a nonperturbative treatment, but is sometimes
less transparent in physical terms since it is not in the usual
continuum language.
In spite of the nonlinear dynamics of
the Liouville theory, a method based on conformal field theory has
now been sufficiently developed to
understand the results of the matrix model
and to offer in some cases a more powerful method in computing
various quantities.
In particular, we can calculate not only partition
functions but also correlation
functions by using the procedure of the analytic
continuation  \NPrefmark{\gouli, \dfku}.
\par
So far only conformal field theories with central charge $ c \le 1$
have been successfully coupled to quantum gravity.
The $c = 1$ model is the richest and the most interesting, and it
is in some sense the most easily soluble.
{}From the viewpoint of the string theory, the $c=1$ model
has at least one (continuous) dimension of target space in which
strings are embedded.
Hence, we can discuss the space-time interpretation in the usual sense
in the $c=1$ model.
Since the Liouville (conformal) mode plays a dynamical role if the
dimensions of the target space is different from the usual critical
dimensions, the theory is called ``noncritical'' string theory.
\par
It has been observed that the $c=1$ quantum gravity can be regarded
effectively as a critical string theory in two
dimensions, since the Liouville field zero
mode provides an additional ``time-like'' dimension besides the
obvious single spatial dimension given by the zero mode of the $c=1$
matter \NPrefmark{\polch}.
We have a physical scalar particle corresponding to the center of mass
motion of the string.
Though it is massless, it is still referred
to as a ``tachyon'' following
the usual terminology borrowed from the critical string theory.
Since there are no transverse directions, the continuous (field)
degrees of freedom are exhausted by the tachyon field.
The partition function for the torus topology
was computed in the Liouville theory, and was found to give precisely
the same partition function as the tachyon field
alone.
However, it has been noted that there exist other
discrete degrees of freedom in the $ c=1$ matter coupled to the 2-D
quantum gravity \NPrefmark{\grklne - \satafact}.
It has been pointed out that the symmetry group relevant to the
dynamics of these discrete states in the $c=1$ quantum gravity is
the area preserving diffeomorphisms whose generators fall into
representations of SU(2) \NPrefmark{\witten}.
Exploiting the SU(2) symmetry, Klebanov and Polyakov
have recently worked out the three point interactions of the
discrete states and have proposed an effective action for these
discrete states \NPrefmark{\klpo}.
\par
This paper is a brief report on our study of the interaction of
tachyons and discrete states in the $c=1$ quantum
gravity \NPrefmark{\masata}.
We have obtained all the possible three point couplings completely
including both tachyons and discrete states by using the operator
product expansion (OPE) of vertex operators.
We have also found that the so-called cocycle factor is needed to obtain
the operator product expansion with the proper analytic behaviour.
\par
%
%
Let us consider the c=1 conformal matter realized by a single bosonic
field $X$ coupled to the 2-D quantum gravity.
After fixing the conformal gauge
$g_{\alpha \beta} = {\rm e}^{\phi} \hat g_{\alpha \beta}$
using the Liouville field $\phi$, the $c=1$ quantum gravity can be
described by the following action  on a surface with a boundary
\NPrefmark{\dika, \seirev, \beku}
$$
\eqalign{
S[\hat g, X, \phi]
= \ & {1 \over 4\pi\alpha'} \int d^2 z \sqrt{\hat g}
\Bigl( \hat g^{\alpha\beta} \partial_\alpha X \partial_\beta X
+ \hat g^{\alpha\beta} \partial_\alpha \phi \partial_\beta \phi
- 2\sqrt{\alpha'} \hat R \phi \cr
& + 4\alpha' \mu \e^{-2 \phi / \sqrt{\alpha'}} \Bigr)
+ {1 \over \pi\sqrt{\alpha'}} \int d \hat s
\left( - \hat k \phi + \sqrt{\alpha'} \lambda
\e^{-\phi / \sqrt{\alpha'}} \right),
}\eqn\action
$$
where $\alpha'$ is the Regge slope parameter, $\hat R$ the scalar
curvature, $\hat k$ the geodesic curvature along the boundary
and $d \hat s$ the line element of the boundaries with respect to
the reference metric $\hat g_{\alpha\beta}$.
We have rescaled the Liouville field $\phi$.
In this paper we will consider only the bulk (or resonant) correlation
functions \NPrefmark{\polyakov}, for which the ``energy'' and
the momentum conjugate to $\phi$ and $X$ respectively are conserved.
For such correlation functions we can use the action without the
cosmological terms by putting $\mu = \lambda = 0$.
\par
There are two types of physical operators.
The open string vertex operators are given by line integrals of
primary fields with boundary conformal weight one along the boundary,
while the closed string vertex operators are given by surface
integrals of primary fields with conformal weight $(1, 1)$.
It is convenient to set $\alpha' = 4 \; (1)$ when we discuss
the closed (open) string vertex operators.
With this convention the integrands of the closed string vertex operators
can be constructed by combining the holomorphic operator and the
anti-holomorphic operator, both of which have the same form as those of
the open string vertex operators.
The holomorphic part of the energy-momentum tensor for $\alpha'=4$
is given by
$$
T(z) = -{1 \over 4} (\partial X)^2 -{1 \over 4} (\partial \phi)^2
- \partial^2\phi.
\eqn\holemtensor
$$
{}From the action, we have
correlation functions of $X$ and $\phi$ for closed string
$$
\VEV{X(z,\bar z) X(w,\bar w)}=\VEV{\phi(z,\bar z) \phi(w,\bar w)}
=-2{\rm ln}|z-w|^2.
\eqn\xcorrfunc
$$
in accord with the convention of Klebanov and
Polyakov \NPrefmark{\klpo}.
\par
Let us first consider the holomorphic part of the vertex operator
corresponding to the open string vertex operators.
They must be a line integral of a primary field
of unit conformal weight.
The simplest field for such operators is the tachyon vertex operator
$$
\Psi^{(\pm)}_p (z) = \e^{ipX(z)} \e^{(\pm p-1)\phi(z)}
\eqn\tachyon
$$
for an arbitrary real momentum $p$.
For higher levels there are non-trivial primary fields
only when the momentum is an integer or a half odd
integer. They are primary fields for
the ``discrete states'' \NPrefmark{\grklne, \polyakov}.
They form SU(2) multiplets and can be constructed
as \NPrefmark{\witten, \klpo}
$$
\Psi^{(\pm)}_{J,m} (z) = \sqrt{(J+m)! \over (2J)! (J-m)!}
\oint {du_{J-m} \over 2\pi i} H_-(u_{J-m}) \cdots
\oint {du_1 \over 2\pi i} H_-(u_1) \Psi^{(\pm)}_{J} (z),
\eqn\discrete
$$
where $J = 0, {1 \over 2}, 1, \cdots \; $; $m= -J, -J+1, \cdots, J$ and
$\Psi^{(\pm)}_{J} (z)$ is the tachyon operator \tachyon\ with
the momentum $p=J$. The integrals are along closed contours surrounding
a point $z$ with $|u_i| > |u_j|$ for $i > j$.
The field $H_-(z)$ corresponds to the lowering operator of the
SU(2) quantum numbers and is one of the SU(2) currents
$$
H_\pm (z) = \e^{\pm i X(z)} = \pm \Psi^{(+)}_{1, \pm1} (z), \quad
H_3 (z) = {1 \over 2} i \partial X(z)
= -{1 \over \sqrt 2} \Psi^{(+)}_{1, 0} (z).
\eqn\sugenerator
$$
The quantum numbers $J$ and $m$ correspond to the ``spin'' and the
magnetic quantum number in SU(2).
Actually, the fields $\Psi^{(\pm)}_{J, m}$ with $m= \pm J$ are not higher
level operators but tachyon operators  \tachyon\ at integer or half odd
integer momenta $\pm J$.
\par
%
In ref.\ \NPrefmark{\klpo} the OPEs of the fields for discrete
states \discrete\ were obtained using the SU(2) symmetry.
Here we make a remark on the analytic property of the OPEs.
The OPE of two vertex operators
gives a  coefficient different in sign depending on the ordering of
the two vertex operators.
Even if we use the radial ordering of the two vertex operators
as usual in conformal field theory,
the OPE is not analytic at $|z| = |w|$.
It is desirable to obtain the analytic OPEs since the
techniques of conformal field theories make full use of the analyticity.
One should multiply the vertex operator \discrete\ by a correction
factor as in the vertex operator construction of the
affine Kac-Moody algebra \NPrefmark{\gool}.
\par
We have succeeded in constructing the necessary correction factor
to recover the analyticity.
After some lengthy argument using the knowledge of integral cubic
lattice, we arrive at the following choice of the cocycle factor
\NPrefmark{\masata}
$$
\varepsilon (\alpha_1, \alpha_2) = (-1)^{2J_1(J_2-m_2-1)} \qquad
 \alpha_i = \sqrt 2 (m_i, J_i-1), \quad i=1,2.
\eqn\cocycleresult
$$
The sign of $J$ in the cocycle factor should be changed according to
the sign of $J$ in the two-vector $\alpha$ corresponding to the
$(-)$ type.
It is easy to see that eq.\ \cocycleresult\ indeed satisfies the cocycle
conditions.
With this cocycle the correction factor is constructed
as \NPrefmark{\gool}
$$
c_\alpha = \sum_{\beta \in \Lambda} \varepsilon (\alpha, \beta)
\ket{\beta} \bra{\beta},
\eqn\calpha
$$
where $\ket{\beta}$ is an eigenstate of the energy and the momentum
with an eigenvalue $\beta$.
Then the corrected operators
$$
\Psi'^{\, (s)}_{J, m} (z) = \Psi^{(s)}_{J, m} (z) c_\alpha, \quad
\alpha = \left\{\eqalign{
\sqrt 2 \, (m, J-1) & \quad {\rm for}\;\, s=+ \cr
\sqrt 2 \, (m, -J-1) & \quad {\rm for}\;\, s=- } \right.
\eqn\cdiscrete
$$
satisfy the OPEs which are analytic in the complex $z$ plane.
\par
We find that after
an appropriate rescaling the corrected operators \cdiscrete\ satisfy
the same OPEs as those given in ref.\ \NPrefmark{\klpo}.
The non-trivial OPEs are given by
$$
\eqalign{
\tilde\Psi'^{\,(+)}_{J_1, m_1}(z) \, \tilde\Psi'^{\,(+)}_{J_2, m_2}(w)
& \sim \, {1 \over z-w} \, (J_2 m_1- J_1 m_2) \,
\tilde\Psi'^{\,(+)}_{J_1+J_2-1, m_1+m_2}(w), \cr
\tilde\Psi'^{\,(+)}_{J_1, m_1}(z) \,
\tilde\Psi'^{\,(-)}_{J_1+J_3-1, -m_1+m_3}(w)
& \sim \, {1 \over z-w} \, (- J_1 m_3 - J_3 m_1) \,
\tilde\Psi'^{\,(-)}_{J_3, m_3}(w).
}\eqn\discreteope
$$
Other OPEs have no singular term.
We have used rescaled fields
$$
\eqalign{
\tilde\Psi'^{\,(+)}_{J,m}(z) &
= \tilde N(J, m) \, \Psi'^{\,(+)}_{J,m}(z), \cr
\tilde\Psi'^{\,(-)}_{J,m}(z) &
= (-1)^{J(2J-1)+J-m} \left[ \tilde N(J, m) \right]^{-1}
\Psi'^{\,(-)}_{J,m}(z),
}\eqn\rescale
$$
$$
\tilde N(J, m) = (2J-1)! \sqrt{J \over 2} N(J, m), \quad
N(J, m) = \left[ {(J+m)!(J-m)! \over (2J-1)!} \right]^{1 \over 2}.
\eqn\njm
$$
\par
%
We shall now generalize these results of the OPEs to include tachyon
operator \tachyon.
We have succeeded to generalize the cocycle operator to the tachyon
case, but we merely refer our paper \NPrefmark{\masata} for the full
account of the construction and write down only the OPE without
the cocycle factors because of lack of space.
{}From the conservation of the energy and the momentum
we find that only four non-trivial OPEs are possible:
$$
\eqalign{
\Psi^{(+)}_{p_1}(z) \, \Psi^{(+)}_{p_2}(w)
& \sim \, {1 \over z-w} \, F^{(+)}_{p_1 p_2} \,
\tilde\Psi^{(-)}_{J_3, 1-J_3}(w) \quad (J_3 = -p_1-p_2+1), \cr
\Psi^{(-)}_{p_1}(z) \, \Psi^{(-)}_{p_2}(w)
& \sim \, {1 \over z-w} \, F^{(-)}_{p_1 p_2} \,
\tilde\Psi^{(-)}_{J_3, J_3-1}(w) \quad (J_3 = p_1+p_2+1), \cr
\tilde\Psi^{(+)}_{J_1, J_1-1}(z) \, \Psi^{(+)}_{p_2}(w)
& \sim \, {1 \over z-w} \, G^{(+)}_{J_1 p_2} \,
\Psi^{(+)}_{p_3}(w) \quad (p_3 = J_1-1+p_2), \cr
\tilde\Psi^{(+)}_{J_1, 1-J_1}(z) \, \Psi^{(-)}_{p_2}(w)
& \sim \, {1 \over z-w} \, G^{(-)}_{J_1 p_2} \,
\Psi^{(-)}_{p_3}(w) \quad (p_3 = 1-J_1+p_2).
}\eqn\tachyonope
$$
The coefficient in the third and fourth OPE in eq.\ \tachyonope\ can be
obtained by using the representation \discrete\ for
$\Psi^{(+)}_{J_1, J_1-1}$ or the similar expression for $m=1-J$ and
directly evaluating the OPE
$$
\eqalign{
G^{(+)}_{J_1 p_2}
& ={\Gamma(1-2p_2) \over 2\Gamma(-2p_3)}
= (-1)^{2J_1-1} \,
{\tilde N(p_3, p_3) \over \tilde N(p_2, p_2)} \, p_2, \cr
G^{(-)}_{J_1 p_2}
& =(-1)^{J_1(2J_1-1)}{\Gamma(1+2p_2) \over 2\Gamma(2p_3)}
= (-1)^{J_1(2J_1-1)} \,
{\tilde N(p_2, p_2) \over \tilde N(p_3, p_3)} \, p_3,
}\eqn\opegco
$$
where $\tilde N(p, p) = {1 \over 2} \Gamma (1+2p)$.
To obtain the coefficient of the first OPE in eq.\ \tachyonope, we
apply the operator $\oint {du \over 2\pi i} H_- (u)$ to both hand sides
of the equation, where the integration contour surrounds both of
$z$ and $w$.
The coefficient of the second OPE in
eq.\ \tachyonope\ can be obtained similarly by applying
$\oint {du \over 2\pi i} H_+ (u)$. We find
$$
\eqalign{
F^{(+)}_{p_1 p_2}
& ={\Gamma(1-2p_1) \over 2\Gamma(2p_2)}
=\left[ \tilde N(p_1, p_1) \tilde N(p_2, p_2) \right]^{-1}
{\pi p_1 p_2 \over 2 \sin (2 \pi p_1)}, \cr
 (-1)^{J_3(2J_3-1)} F^{(-)}_{p_1 p_2}
& = - {\Gamma(1+2p_2) \over 2\Gamma(-2p_1)}
= {2 \over \pi} \,
\tilde N(p_1, p_1) \tilde N(p_2, p_2) \sin (2\pi p_1).
}\eqn\opefco
$$
\par
%
The coefficients of the OPE determine the three-point correlation
functions
of the physical operators, which can be summarized by the effective
action.
Introducing a variable $g^{(s)}_{J,m}$ $(s=\pm)$ for each discrete
state,
the cubic terms of the effective action for discrete states determined
by the OPEs \discreteope\ are \NPrefmark{\klpo}
$$
S_3 = {1 \over 2} \sum_{J_1, m_1, J_2, m_2} (J_2 m_1-J_1 m_2)
f^{ABC} g^{(-) A}_{J_1+J_2-1, -m_1-m_2}
g^{(+) B}_{J_1, m_1} g^{(+) C}_{J_2, m_2}
\int d\phi,
\eqn\cubic
$$
where we have introduced the Chan-Paton index $A$ in the adjoint
representation of some Lie algebra and have factored out the Liouville
volume $\int d\phi$.
\par
In ref.\ \NPrefmark{\klpo} it was shown that the terms in the
cubic interaction \cubic\ which depend only on the integer modes
$g^{(s) A}_{J, m} \;\; (J, \, m \in \Z)$ can be written in a
compact form by introducing a scalar field on $\R \times {\rm S}^2$
$$
\Phi_0 (\phi, \theta, \varphi)
= \sum_{s, A, J, m} T^A g_{J, m}^{(s) A} M^s(J, m)
D^J_{m 0} (\varphi, \theta, 0) \e^{(s J-1)\phi}.
\eqn\integermode
$$
Here, $T^A$ are the representation matrices of the Lie algebra,
$D^J_{m 0}$ are components of the SU(2) rotation
matrix \NPrefmark{\rose} and $M^s (J, m)$ are  the normalization factor
$$
D^J_{m m'} (\varphi, \theta, \psi)
= \bigl\langle J m \bigr\vert \e^{-i\varphi J_z} \e^{-i\theta J_y}
\e^{-i\psi J_z} \, \bigl\vert \, J m' \bigr\rangle,
\eqn\rmatrix
$$
$$
M^+ (J, m) = {N(J, m) N(J, 0) \over J}, \quad
M^- (J, m) = {(-1)^m \over 4\pi} {J(2J+1) \over N(J, m) N(J, 0)}.
\eqn\integermm
$$
The effective action can be written
in terms of the field $\Phi_0$ using $x^i = (\theta, \varphi)$
$$
S_3^{(1)} = \int d\phi \e^{2\phi} \int\nolimits_{S^2} d\theta
d\varphi \,
\epsilon^{ij} \, {\rm Tr} \left( \Phi_0
{\partial \Phi_0 \over \partial x^i}
{\partial \Phi_0 \over \partial x^j} \right) .
\eqn\integeraction
$$
\par
We have succeeded to generalize this construction to the terms
containing half odd integer modes as well as integer modes.
We introduce two spinor fields
$\Phi_{1 \over 2}$ and $\Phi_{-{1 \over 2}}$ on
$\R \times {\rm S}^2$ for half odd integer modes
$g_{J, m}^{(s) A} \;\; (J, \, m \in \Z + {1 \over 2})$
$$
\Phi_\mu (\phi, \theta, \varphi)
= \sum_{s, A, J, m} T^A g_{J, m}^{(s) A} M^s_\mu(J, m)
D^J_{m \mu} (\varphi, \theta, 0) \e^{(s J-1)\phi}
\quad \left( \mu = \pm {1 \over 2} \right),
\eqn\halfintegermode
$$
where
$$
M^+_\mu (J, m) = {N(J, m) N(J, {1 \over 2}) \over J+{1 \over 2}}, \quad
M^-_\mu (J, m) = {(-1)^{m+\mu} \over 4\pi} {2J(J+1)
\over N(J, m) N(J, {1 \over 2})}.
\eqn\halfintegermm
$$
Note that $\Phi_{1 \over 2}$ and $\Phi_{-{1 \over 2}}$ have the same
coefficients $g_{J, m}^{(s) A}$ and are not independent.
In order to write down the effective action in terms of these fields
we need covariant derivatives on ${\rm S}^2$ acting on spinor fields
$\Phi_{\mu}$. They are given by
$$
\nabla_\pm = \mp \partial_\theta -
{1 \over \sin\theta} (i \partial_\varphi - \mu \cos\theta)
\eqn\covd
$$
when acting on $\Phi_\mu$.
The effective action can be written as
$$
S_3^{(2)} = \int d\phi \e^{2\phi} \int\nolimits_{S^2} d\theta d\varphi
\sin\theta \; {\rm Tr} \left( \Phi_0 \left[ \nabla_+
\Phi_{-{1 \over 2}}, \nabla_- \Phi_{1 \over 2} \right] \right).
\eqn\halfintegeraction
$$
The sum of eqs.\ \integeraction\ and \halfintegeraction\ gives the
complete cubic terms for the discrete states \cubic.
\par
Apart from the special case of the compact boson $X$ with the self-dual
radius, we have tachyons with momenta other than integer or half odd
integer which should be included in the effective action.
The OPE results \tachyonope\ can be summarized as two types of terms
in the effective action involving tachyons: two tachyons with the same
chirality $(+)$ or $(-)$ couple to the single discrete state
of the $(+)$ type.
We have succeeded to write down the local effective action involving
tachyons, for which we refer our paper \NPrefmark{\masata}.
\refout
\end
, \cr
& \qquad\qquad\qquad\qquad\qquad\qquad (J_1 < J_3+1)